\documentclass[11pt]{article}

\usepackage{amsmath}
\usepackage{amsthm}
\usepackage{amssymb}
\usepackage{amscd}
\usepackage{epsf}
\usepackage{rotating}
\usepackage{wrapfig}

\chardef\bslash=`\\ 

\makeatletter
\def\verbatim{\interlinepenalty\@M \@verbatim
  \leftskip\@totalleftmargin\advance\leftskip2pc
  \frenchspacing\@vobeyspaces \@xverbatim}
\makeatother
\theoremstyle{plain}

\theoremstyle{remark}

\numberwithin{equation}{part}

\def\1I{\relax{\rm 1\kern-.25em \rm l}} 

\newcommand{\vqm}{V_{\rm QM}}
\newcommand{\jost}{{\cal J}}

\def\Rahmen#1#2#3 {
   \vbox{\hrule height#2
         \hbox{
               \vrule width#2
               \hskip#1
               \vbox{
                     \vskip#1{}
                     \hbox{#3}
                     \vskip#1
                    }%
               \hskip#1
               \vrule width#2
              }
         \hrule height#2
        }}
\def\href#1#2{#2}

\begin{document}

\thispagestyle{empty}
\rightline{hep-th/yymmnn}
\vspace{2truecm}
\centerline{\bf \Large Note on the holographic $c$-function}
\vspace{.5truecm}

\newcounter{Institut}
\vspace{1.5truecm}
\centerline{\bf Andr\'e Miemiec}

\vskip1cm
\parbox{.9\textwidth}{
\centerline{Department of Theoretical Physics}
\centerline{University of Turin}
\centerline{INFN-Turin}
\centerline{Via P.Giuria 1}
\centerline{10125 Torino}
\centerline{Italia}
\centerline{miemiec@to.infn.it}
}

\vspace{.4truecm}

\vspace{1.0truecm}
\begin{abstract}
 \noindent
We discuss the holographic c-function and describe an algorithm 
for the practical computation of the changing central charge in 
arbitrary RG-flows. The conclusions are drawn from studying a 
particular example, which is worked out in detail. The 
renormalisation procedure of hep-th/0112150 necessary to obtain the 
central charge is reviewed.    
\end{abstract}
\bigskip \bigskip
\newpage


\noindent
Since the work of Zamolodchikov \cite{Zamolodchikov:1986gt}, the proof 
of a $c$-theorem for quantum field theories in more than two dimensions has
remained an open problem. In the last years the idea of holography, 
namely the extension of the AdS/CFT correspondence to non conformal theories, 
has provided new tools for addressing this issue. In 
\cite{Girardello:1998pd,Freedman:1999gp} a holographic $c$-theorem has 
been established. It is defined for quantum field theories, which allow a 
holographic description in terms of gravity. This is quite a restriction.
In fact the existence of this construction seems to hinge on a special 
identity fulfilled for the coefficients $a$ and $c$ of the two contributions 
to the conformal anomaly, i.e. the Euler density and the Weyl 
tensor \cite{Anselmi:1997am}. For a holographic renormalisation group 
flow (RG-flow) these coefficients 
coincide. Within field theory there are counterexamples for $c$-theorem 
conjectures based on c, while for all known field theory examples 
$a_\mathrm{UV} \geq a_\mathrm{IR}$ for the coefficient of the Euler density. So the 
$c$-theorem should be actually dubbed $a$-theorem.\\
Inspired by \cite{Anselmi:2000fu} we constructed explicit examples of 
holographic RG-flows between two conformal field theories (CFT) by glueing 
asymptotically AdS-spacetimes together 
and studied the corresponding $c$-functions \cite{Martelli:2001aa,Miemiec}. 
A crucial feature of the examples was the necessity to introduce 
artificially a ``renormalised'' $c$-function in order to obtain the 
expected values for the central charges. The renormalisation must be 
performed in order to get rid of a divergence caused by the presence 
of a ``massless pole'', which lacks a physical interpretation. We 
argued that the appearance of the divergence was an artefact of the 
construction, i.e. induced by the  glueing procedure, which required only 
continuity at the gluing point. We speculated that 
for a smooth RG-flow no renormalisation must be performed. Here we  
want to push the topic a little bit further by analysing a smooth example. 
We first shortly summarise the construction of the c-function and then 
focus on the lesson taught by the smooth model. \\

\noindent
In \cite{Anselmi:2000fu} a natural definition of a holographic $c$-function 
was proposed, the so-called ``canonical'' $c$-function,
which is related to the OPE of the stress-energy tensor \cite{Anselmi:1998rd} 
as (here in $d=4$)
\begin{eqnarray}\label{TTcorr}
\langle T_{\mu\nu}(x)T_{\rho\sigma}(0) \rangle & = &
-\frac{1}{48\pi^4}\Pi^{(2)}_{\mu\nu\rho\sigma}
\left[\frac{c(x)}{x^4}\right]+\pi_{\mu\nu}
\pi_{\rho\sigma}\left[\frac{f(x)}{x^4}\right] 
\end{eqnarray}
where $\pi_{\mu\nu}=\partial_\mu\partial_\nu-\eta_{\mu\nu}\Box$ and
$\Pi^{(2)}_{\mu\nu\rho\sigma}=2\pi_{\mu\nu}\pi_{\rho\sigma}-
3(\pi_{\mu\rho}\pi_{\nu\sigma}+\pi_{\mu\sigma}\pi_{\nu\rho})$.
For a 4-d conformal QFT the function $f(x)$ vanishes and $c(x)$ becomes 
a constant, the so called central charge $c$.   
We want to compute this 4-dimensional correlator by applying the 
AdS/CFT correspondence to the 5-dimensional gravity action below: 
\begin{eqnarray}\label{action}
  S &=& \int d^{d+1}x\,\sqrt{g}\, 
                     \left[\vbox{\vspace{2ex}}\right. 
                            R -\frac{1}{2} (\partial\phi)^2-V(\phi)
                     \left.\vbox{\vspace{2ex}}\right] 
    - 2\int d^d x \sqrt{\hat{g}}\,K
\end{eqnarray}
This is done as follows. First one has to choose a classical solution
\begin{eqnarray}\label{metrik}
  ds^2 & = & e^{2A(z)}\,(\,dz^2 \,+\, \eta_{ij}\,dx^i dx^j\,),
  \qquad
  \phi    ~ = ~ \phi (z)
\end{eqnarray} 
of the equation of motion of the action (\ref{action})  
\begin{eqnarray}\label{EOM}
   \frac{d A}{d z} ~=~  \frac{e^{A}}{2(d-1)}\,W~,\qquad
   \frac{d \phi}{d z} ~=~  -\,e^{A}\,W_\phi
\end{eqnarray}
where as $z$ moves from zero to infinity, $\phi(z)$ interpolates between  
two nearby extrema of the potential $V(\phi)$. Needless to mention that 
we have to choose an appropriate  $V(\phi)$, of course. If $\phi(z)$ 
approaches one of these extrema the geometry of the metric asymptotes to AdS.
The function $W(\phi)$ appearing above is related to the potential by 
\begin{eqnarray}
        V(\phi)\,=\,\frac{1}{2}\,W'{}^2-\frac{d}{4(d-1)}\,W^2~.
\end{eqnarray} 
The gravity calculation of $<TT>$-correlator for 
a specified background flow was done in \cite{Gubser:1998bc, witten,Brandhuber:1999hb} and 
boils down to the solution of the fluctuation equation for the linearised 
transverse traceless graviton (TT)  
\begin{eqnarray}
   h_{ij}^{TT} ~=~ e^{ikx}\,\chi(z)\,\xi_{ij} (x)~, 
\end{eqnarray}
which satisfies the equation 
\begin{eqnarray}\label{schr1}
    \left[\,
             -\,\frac{d^2}{dz^2} ~+~ 
              \left(\,\vqm (z)\,-\,k^2\,\right)\,
    \right]\,e^{\frac{d-1}{2}A(z)}\chi(z)
    & = & 0~\\[1.5ex]
   \vqm  ~=~  \left(\frac{d-1}{2}\right)^2\hspace{-1ex} A'{}^2 ~+~
              \frac{d-1}{2}\,A''~.&&
\end{eqnarray}
The $c$-function in eq. (\ref{TTcorr}) can be obtained via 
the four dimensional Fourier transform of the so called flux 
factor ${\cal F}(q)$ (~$q\,=\,i\,|k|$~), which can be constructed 
out of the fluctuation $\chi(z)$:  
\begin{eqnarray}\label{fourier}
 \frac{c(z)}{z^4} 
    &=& {\mathfrak{F}}
        \left[\,
                \frac{{\cal F}(q)}{q^4}\,
        \right](z)
\end{eqnarray}
with 
\begin{eqnarray}\label{flux}
 {\cal F}(q) &=& \frac{1}{\varepsilon^{2\lambda-1}}\,
                          \lim\limits_{z\rightarrow\varepsilon}\,
                       \frac{d}{dz}\,\frac{\chi(z)}{\chi(\varepsilon)}~.
\end{eqnarray}
The evaluation of eq.~(\ref{flux}) for a generic potential is the one 
part of the problem. We solved it by showing that the potentials which 
arise from a given background are those to which scattering theory applies. 
The solution $\chi(z)$ of the fluctuation equation one has to choose is 
fixed by regularity in the IR. This is due to an finite action argument. 
Having chosen this solution, which is unique up to a scaling, one simply 
considers its decomposition into two solutions with prescribed power law 
close to the origin. The solution reads\footnote
{
  For technical details see appendix \ref{AppNotation}.
}
\begin{eqnarray}
   \chi(q,z) 
   = e^{-\frac{d-1}{2}\,A(z)}\,
     \left[\,
              \jost( \lambda,-q)\,\varphi_{\rm irr}(z) - 
              \jost(-\lambda,-q)\,\varphi_{\rm reg}(z)\,
     \right]
\end{eqnarray} 
where $\jost(\pm\lambda,-q)$ are the so called Jost functions. Here the 
Jost functions contain the necessary informations about the potential.    
The flux computed in \cite{Martelli:2001aa} reads 
\begin{eqnarray}\label{firstresult}
{\cal F}(q) & = &  {\cal F}_\mathrm{log}(q) ~+~  {\cal F}_\mathrm{int}(q)
\end{eqnarray}      
with 
\begin{eqnarray}
 {\cal F}_\mathrm{log}(q) & = & \,\frac{(-1)^{\lambda+1}}{2^{2\lambda-2}[\Gamma(\lambda)]^2}\,q^{2\lambda}\,\log\,q \\
 {\cal F}_\mathrm{int}(q) & = &                            
                               -\,2\,\lambda\, R_\mathrm{UV}^{2\lambda-1}\,
                                \frac{\jost( -\lambda,-q)}{\jost(\lambda,-q)}\label{FInt}~.
\end{eqnarray}
In \cite{Martelli:2001aa,Miemiec} we observed the necessity to accomplish 
the definition of the $c$-function by a subtraction procedure in order to
obtain the expected values for the central charges. The reason for this 
``renormalisation'' could not be explained. Relying on a heuristic argument,
we conjectured that the subtraction should be absent for a continuous flow.\\

\label{largeqlimit}
\noindent
The Jost functions $\jost( -\lambda,-q)$ vanishes for large $q$. So the 
term ${\cal F}_\mathrm{log}(q)$ is the only one, which survives in the 
large $q$-limit. It determines the central charge $c_\mathrm{UV}$ in the 
UV, which can be conveniently normalised by multiplying the whole flux factor 
with an appropriate constant. A practical choice is to set 
$c_\mathrm{UV}\,=\,1$.
The exact computation of ${\cal F}_\mathrm{int}(q)$ turns out to be rather 
difficult. The problem is easy to understand. We are interested in the small
$q$-expansion of ${\cal F}_\mathrm{int}(q)$ and in particular in the
logarithmic terms, which vanish close to zero. The $q$-expansion contains 
terms divergent for small $q$, too. So we have to extract a subleading order
numerically.\\ 

\noindent
To make the task tractable, we decided to map the problem 
into an 2-point boundary value problem, which is numerically accessible. 
This is achieved by first rescaling the solution according to 
$\tilde{y}(z)\,=\,z^{3/2}\cdot y(z)$ so 
that close to $z\,=\,0$ the irregular solution approaches the value 1, 
while the regular solution tends to zero as before. Any solution must be 
decomposable into a sum of the irregular and regular solutions. Taking this  
and the non uniqueness of $\chi(z)$ due to a rescaling (~see eq.~(\ref{flux})~) into account one may impose at the left boundary the 
condition $y(0)\,=\,1$. Mapping now the interval $I_z\,=\,[\,0,\,\infty\,]$ 
by means of the transformation $z=-1+\frac{1}{\sqrt{x}}$  to the interval 
$I_x\,=\,[\,0,\,1\,]$ one can single out the exponentially decaying solution 
by the additional boundary condition $\tilde{y}(0)\,=\,0$. Thus we obtain
\footnote
%
{
  For the definition of $U$ see eq.~(\ref{ScatteringPotential}).
},
\begin{eqnarray*}
 -4\,x^3\,\frac{d^2\tilde{y}(x)}{dx^2}\,-\,6\,x^2\,\,\left(\,1+\frac{1}{1-\sqrt{x}}\,\right)\,\frac{d\tilde{y}(x)}{dx}&+&\left(\,U(x)+q^2\,\right)\,\tilde{y}(x) ~=~ 0\\[2ex]
 {\rm BC}:\hspace{3ex}\tilde{y}(0)\,=\,0,\hspace{3ex}\tilde{y}(1)&=&1\,,
\end{eqnarray*}
the numerical solution of which is performed via a finite difference 
algorithm \cite{FDiff}. Please note that the contraction of the interval 
provides us with a substantial benefit. The equal spacing of the contracted
interval means an unequal spacing on the positive line, i.e. close to zero 
we collect more and more mesh points.
The two alternatives, the shooting algorithm and the solution via an integral 
equation, did not work. The first due to the peculiar boundary conditions we 
have to impose and the other since the numerical data lack precision we must 
obtain.\\

\noindent
Now we are going to investigate a particular example of a smooth RG-flow 
defined by the quantum mechanical potential 
\begin{eqnarray*}
     \vqm(z) &=& \frac{15}{4}\,\frac{1}{z^2} \,+\, U(z)
             ~=~ \frac{15}{4}\,\frac{1}{z^2} \,+\, \frac{z}{(1+z^2)^2}\, ,
\end{eqnarray*}
which meets all the requirements imposed in \cite{Martelli:2001aa}, i.e. 
it is dominated at zero and infinity by the angular momentum barrier and 
smooth everywhere else in between. So it corresponds to an 
interpolating RG-flow. 
The result of the computation of $F_\mathrm{int}(q)/q^2$ is shown in 
Fig.~\ref{figureFModell1} below.\\
\parbox{\textwidth}{\vspace{0.2cm}
\parbox{0.4\textwidth}{\vspace{0.2cm}
  \refstepcounter{figure}
  \label{figureFModell1}
  \begin{center}
  \begin{turn}{-90}
  \makebox[4cm]{
     \epsfxsize=4cm
     \epsfysize=4cm
     \epsfbox{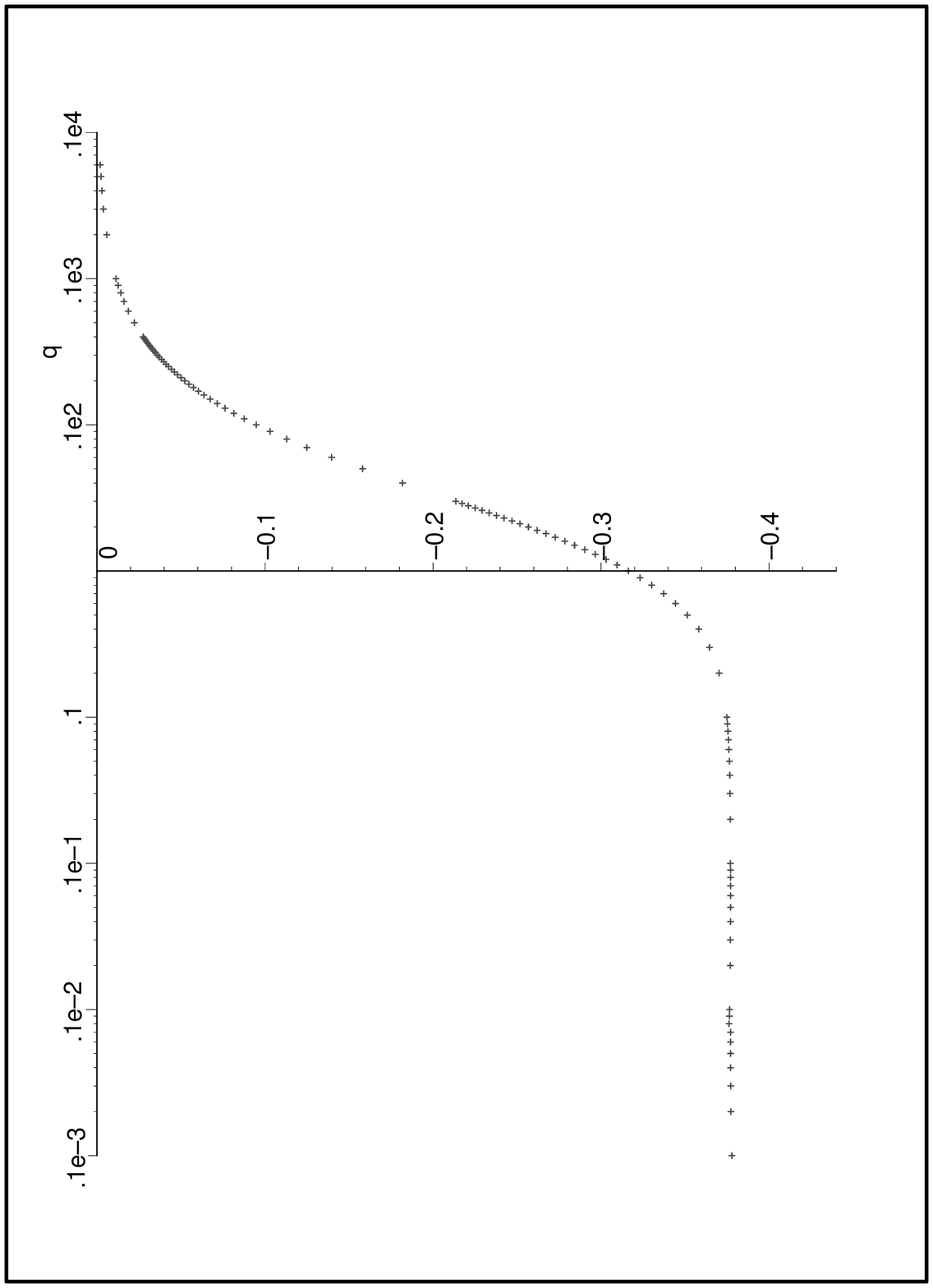}
  }
  \end{turn}
  \end{center}
  \center{{\bf Fig.{\thefigure}.} $\mathcal{F}_\mathrm{int}(q)/q^2$}
}\hspace{1cm}
%
\parbox{0.4\textwidth}{\vspace{0.2cm}
  \refstepcounter{figure}
  \label{figureBoth}
  \begin{center}
  \begin{turn}{-90}
  \makebox[4cm]{
     \epsfxsize=4cm
     \epsfysize=4cm
     \epsfbox{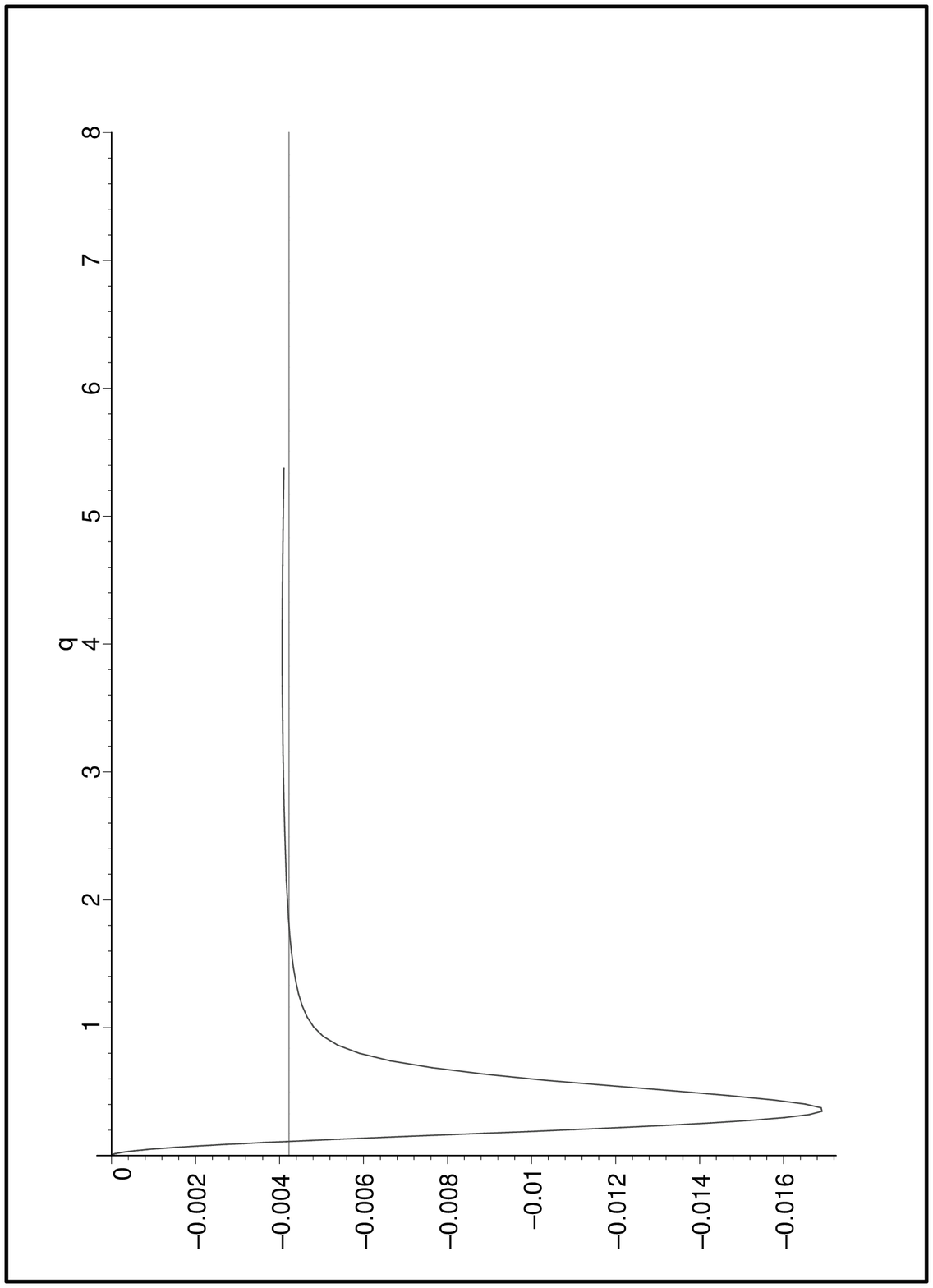}
  }
  \end{turn}
  \end{center}
  \center{{\bf Fig.{\thefigure}.} $c^\mathrm{int}_\mathrm{ren}(z)$}
}\hfill
%
}\\[2.5ex]
The only cosmetic correction we have done here is to subtract the constant 
appearing in the $q$ expansion of $F_\mathrm{int}(q)$\footnote
{
  The value of the constant is $-0.598164897814$
}.
It is important to notice that for very small $q$ the remaining function 
becomes a constant. Again this signals the presence of a $q^2$-term in the 
expansion of $F_\mathrm{int}(q)$, dubbed ``massless pole'' in the 
introduction. 
In order to obtain the proper behaviour of a $c$-function, we have to perform 
the renormalisation procedure introduced in \cite{Martelli:2001aa} again, i.e. 
we must subtract the divergence introduced by the ``massless pole''\footnote
{
  One must subtract $0.09428\cdot x^2$
}. 
The Fourier transformation of eq.~(\ref{fourier}) is performed using a code 
provided in \cite{fht}. In Fig.~\ref{figureBoth} we show the result 
of the Fourier transformation after applying the renormalisation. 
The actual computation of Fig.~\ref{figureBoth} involves an additional trick 
because the numerical Fourier transformation is not stable. Instead the 
plot is 
generated by transforming $F_\mathrm{int}(q)$ multiplied by a gaussian 
distribution. This is merely a sort of cutoff and does not 
spoil the validity of the result as long as we are only interested in the
behaviour at large $q$. This is due to the fact that the Fourier 
transformation just intertwines the region close to zero in $q$ with far 
from zero in $x$ and vice versa. The Gaussian does not affect the behaviour 
of $F_\mathrm{int}(q)$  for small $q$. \\
As a cross check for the validity of the content of Fig.~\ref{figureBoth} 
for large $x$ we just performed the inverse Fourier transformation on the 
unrenormalised data underlying Fig.~\ref{figureBoth} in order to see if we 
can reconstruct Fig.~\ref{figureFModell1} correctly. The ratio of the 
original data shown in Fig.~\ref{figureFModell1} to the reconstructed ones 
is shown in Fig.~\ref{figureRatio}.\\  
\parbox{\textwidth}{\vspace{0.2cm}
  \refstepcounter{figure}
  \label{figureRatio}
  \begin{center}
  \begin{turn}{-90}
  \makebox[4cm]{
     \epsfxsize=4cm
     \epsfysize=4cm
     \epsfbox{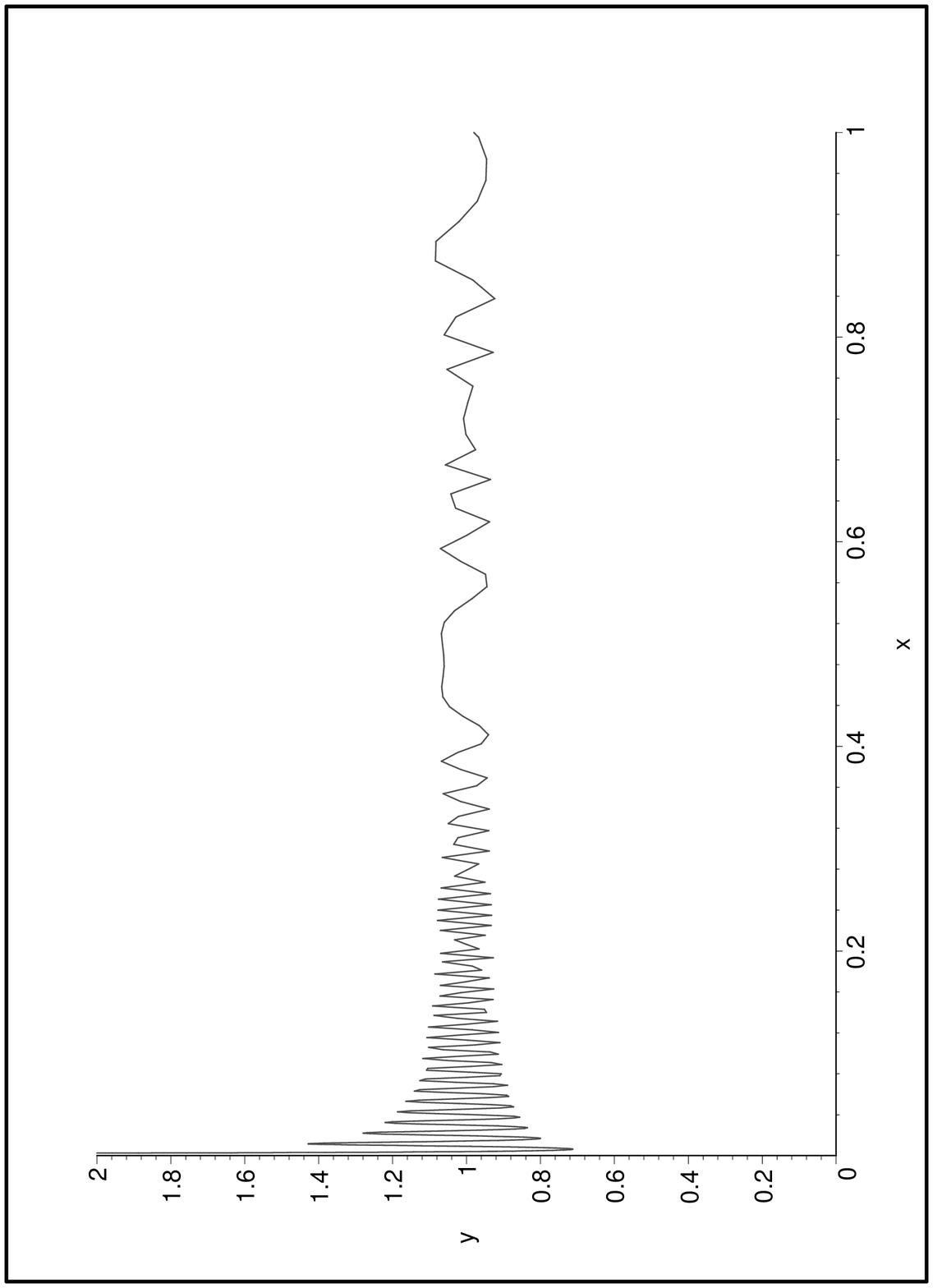}
  }
  \end{turn}
  \end{center}
  \center{{\bf Fig.{\thefigure}.} Ratio $
    r\,=\,\frac{{\mathcal{F}}_{\mathrm{int}}(q)}{{\mathcal{F}}^{\rm b}_{\mathrm{int}}(q)}$ }
}\\[2.5ex]
%
The ratio is close to one nearly everywhere with exception of $q$ very close
to zero. This is not a blow for our procedure because the deviation is related
to the fact that we have not tabulated the transformed function on the whole 
real line but just on a finite interval.\\

\noindent
The sharp peak in Fig.~\ref{figureBoth} is a complete artefact of the 
multiplication by the Gaussian hence completely meaningless. The 
straight line in the same figure denotes the value which must be approached 
for $x$ very large in order to represent a decline in the central charge 
to 2/3 of the original value. With a different sampling of the data it is 
possible to find the right behaviour of the Fourier transform for small 
$x$, too.
The result is shown in Fig.~\ref{figureCModell1}. Here we have summed up 
the two contributions to the c-function coming from 
${\mathcal{F}}_{\mathrm{log}}(q)$ and ${\mathcal{F}}_{\mathrm{int}}(q)$ 
and rescaled to make the UV-charge becoming $1$.  \\
\parbox{\textwidth}{\vspace{0.2cm}
  \refstepcounter{figure}
  \label{figureCModell1}
  \begin{center}
  \begin{turn}{-90}
  \makebox[4cm]{
     \epsfxsize=4cm
     \epsfysize=4cm
     \epsfbox{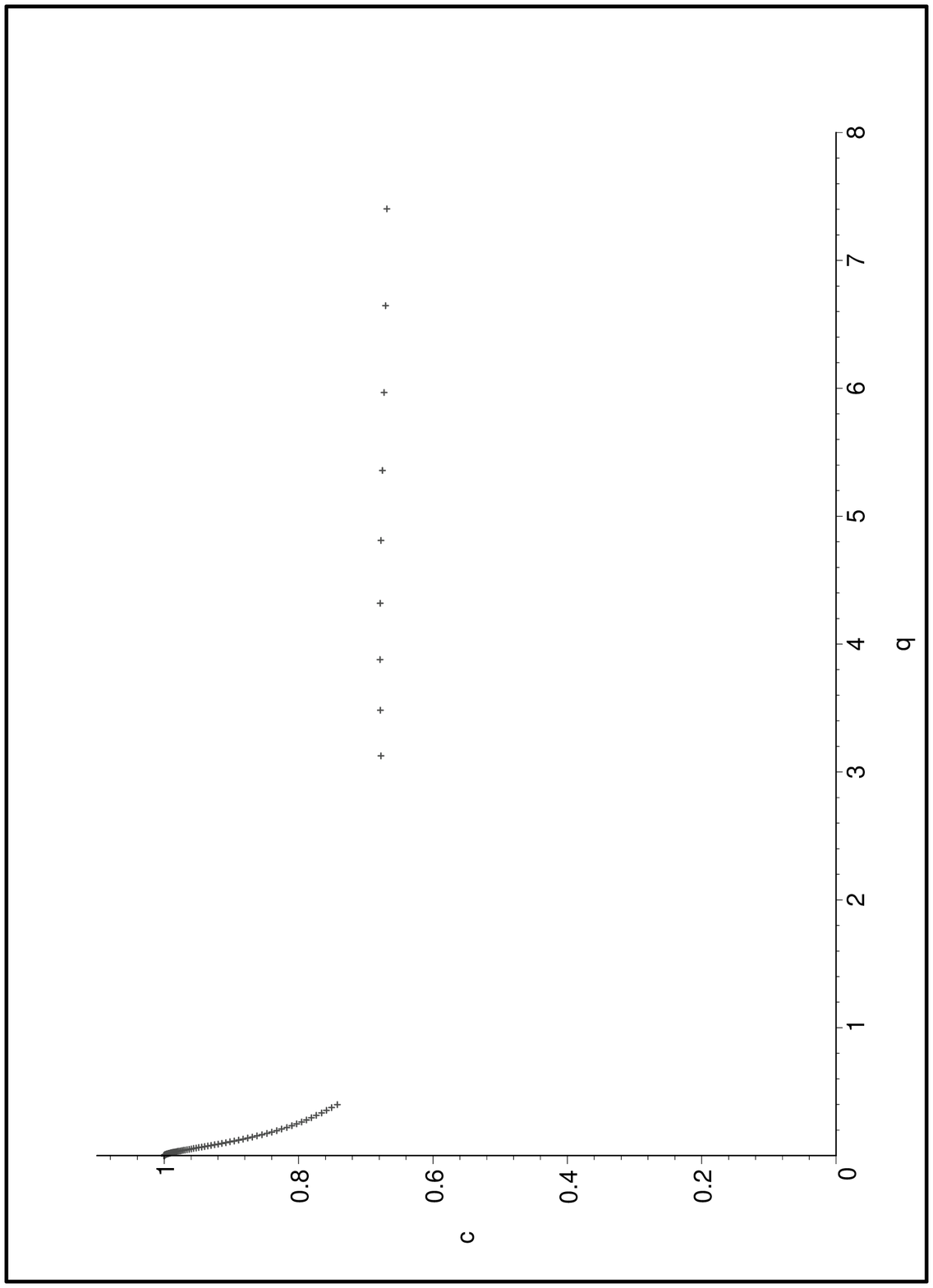}
  }
  \end{turn}
  \end{center}
  \center{{\bf Fig.{\thefigure}.} $c_{\rm ren}(z)$}
}\\[2.5ex]
%
As a comparison we computed the gravitational $c$-function of 
\cite{Girardello:1998pd,Freedman:1999gp}. In the last two papers  
a holographic $c$-theorem was established using the warp factor $A(z)$ 
appearing in the metric eq.~(\ref{metrik}). It is used to introduce a 
function, 
\begin{eqnarray}\label{grav_c}
   c(z) &=& -\,\frac{e^{(d-1)A(z)}}{A'(z)^{d-1}}~, 
\end{eqnarray}
which satisfies all the properties of a $c$-function, i.e. $c(z)$ 
is a monotonous function, which correctly interpolates between the UV and IR 
central charges. The factors in the definition eq.~(\ref{grav_c}) are 
chosen in order to ensure $c_\mathrm{UV}\,=\,1$. So by construction the two 
$c$-functions, eq.~(\ref{fourier}) and eq.~(\ref{grav_c}), can be made to 
become equal at $z\,=\,0$ (cf. second paragraph after eq.~(\ref{FInt}) on 
page \pageref{largeqlimit}).
By a straightforward calculation we express the suitable normalised 
$c$-function entirely in terms of the zero mode 
$\psi_0(z)\,=\,e^{\frac{d-1}{2}A(z)}$ of eq.~(\ref{schr1}). It reads 
\begin{eqnarray}\label{C_Grav}
     c_{\rm grav}(z) &=& -\,\left(\,\frac{d-1}{2}\,\right)^{d-1}\,\frac{\psi_0(z)^{d+1}}{\psi_0'(z)^{d-1}}~. 
\end{eqnarray}
For $d\,=\,4$ we can compute the gravitational $c$-function by constructing 
the zero mode of eq.~(\ref{schr1}) applying the finite difference algorithm 
once more. The corresponding plot is shown in Fig.~\ref{figureC_Grav_Modell1} 
below. The horizontal line again represents the value 2/3. The both 
$c$-functions coincide in their predictions for the central charge. \\
\parbox{\textwidth}{\vspace{0.2cm}
  \refstepcounter{figure}
  \label{figureC_Grav_Modell1}
  \begin{center}
  \begin{turn}{-90}
  \makebox[4cm]{
     \epsfxsize=4cm
     \epsfysize=4cm
     \epsfbox{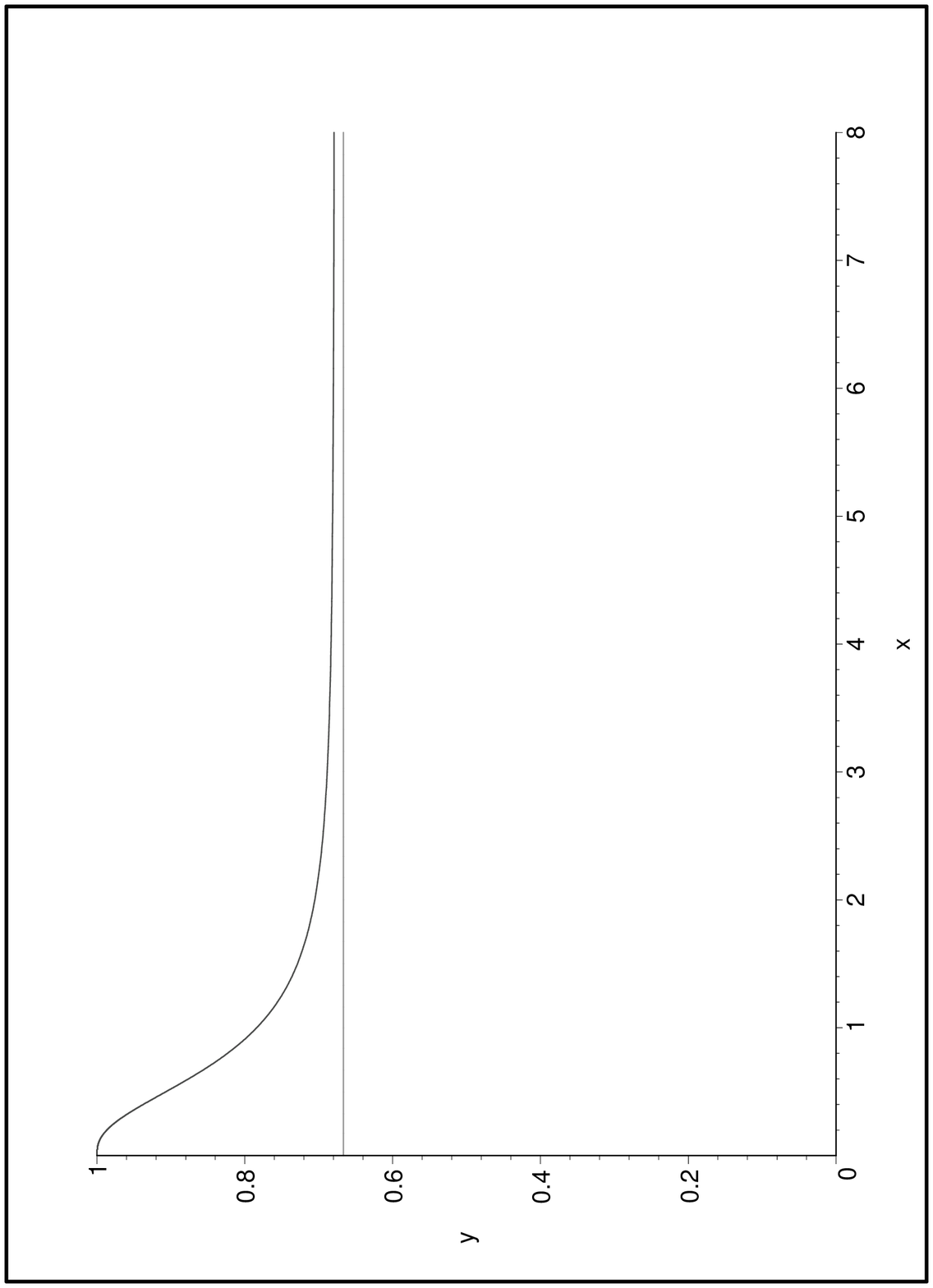}
  }
  \end{turn}
  \end{center}
  \center{{\bf Fig.{\thefigure}.} $c_{\rm grav}(z)$}
}\\[2.5ex]
%

\section*{Conclusions}

We considered the holographic $c$-function connected with a RG-flow, which 
can be computed from a supersymmetric quantum mechanical scattering problem.
The purpose of this work was to show that for a smooth scattering potential 
the $c$-function does not require the renormalisation procedure introduced 
in \cite{Martelli:2001aa}. Instead we found that even here a ``massless 
pole'' appears and renormalisation must be applied again.\\
The problem of the ``massless pole'' is related to the following fact. 
For $\frac{c(x)}{x^4}$ to have a well-defined Fourier transform, it has to
be of the form
\begin{equation}
\frac{c(x)}{x^4} = \partial^2 \left (\frac{1}{x^2} h(t) \right)\, ,\quad
t=\ln \mu^2 x^2 \, ,
\end{equation}
with some suitable function $h(t)$ \cite{Anselmi:1997am}. 
The massless pole is of a peculiar type with respect to the structure of h(t). 
This is due to the fact that the flux ${\mathcal{F}}(q)$ can be expressed as
\begin{eqnarray}
                  {\mathcal{F}}(q) \,=\, 
                     q^5\,\cdot \int\limits_0^\infty h(t) J_1(qx) dx
\end{eqnarray}
and each $t^k$ transforms into terms of the type $q^4\cdot P_k(\ln q)$ with 
$P_k$ a polynomial of degree $k$. The highest power in $P_k(\ln(q))$ is 
$(-1)^k2^k (\ln q)^k$. Unless there is an astonishing conspiracy between 
all the powers $t^k$ no isolated $q^2$ can be obtained.\\ 
The final goal is a proof of a holographic $c$-theorem based entirely on 
the field theoretic definition given by eq.~(\ref{TTcorr}). The example 
points out that the definition of the flux-factor must be improved or 
accompanied by the renormalisation procedure in order to get rid of the 
divergence, which apparently  has no physical meaning.   
The possibility to express the gravitational $c$-function completely in 
terms of the zero mode is interesting for itself and constitute the link 
of both definitions. This will be investigated more carefully in the future.

\vskip1cm
\noindent
{\bf Acknowledgement:} The work of the author is supported by the European 
Commission RTN programme HPRN-CT-2000-00131.\\

\begin{appendix}

\section{Notation}
\label{AppNotation}

The two independent solutions $\varphi_{\rm reg}(z)$ and 
$\varphi_{\rm irr}(z)$ of
\begin{eqnarray*}
    \left[\;
             -\,\frac{hd^2}{dz^2} ~+~ \left(\,\vqm ~-~ k^2\,\right)\;
    \right]\,\varphi(z) ~=~ 0
\end{eqnarray*}
with 
\begin{eqnarray}\label{ScatteringPotential}
        \vqm(z) &=& \frac{\lambda^2-1/4}{z^2} ~+~ U(z)
\end{eqnarray}
and $\lambda\,=\,\frac{d}{2}$ are distinguished by their power 
behaviour at $z\,=\,0$. The regular solution is uniquely defined by 
the asymptotic behaviour at zero.
\begin{eqnarray*}
       \varphi_{\rm reg}(z) &=& z^{\lambda+1/2}\cdot\left(\,1~+~{\mathcal{O}}(1)\,\right)
\end{eqnarray*}
Instead the irregular solution is not uniquely defined by the asymptotic 
behaviour at zero. One can add an arbitrary multiple of 
$\varphi_{\rm reg}(z)$ without changing the asymptotics. We choose the
following combination:
\begin{eqnarray*}
       \varphi_{\rm irr}(z) &=&
       z^{-\lambda+1/2}\cdot\left(\,1~+~{\mathcal{O}}(1)\,\right) 
      ~+~ \frac{(-1)^{\lambda+1}}
               {2^{2\lambda-1}\Gamma(\lambda)\Gamma(\lambda+1)}\cdot 
               q^{2\lambda}\cdot
               \left[\,
                        \ln\left(qx\right) \,+\, {\rm const}\,
               \right]\cdot\varphi_{\rm reg}(z)
\end{eqnarray*}
The value of the dummy ``const'' depends on $\lambda$ and for $\lambda\,=\,2$ 
it becomes $\ln 2+\gamma-3/4$. It is chosen in order to reproduce the 
$K_2(qx)$-Bessel function in the case $U(z)\,=\,0$. The term $\ln q$ is 
responsible for the splitting of ${\mathcal F}(q)$ in the sum of two terms 
(~see eq.~\ref{firstresult}~). The dummy ``const'' can't  be used to 
kill the ``massless pole''.

\end{appendix}


\end{document}